\documentstyle[pre,aps,epsfig,multicol]{revtex}
\begin{document}
\title{Thin front propagation in steady and unsteady cellular flows}
\author{M. Cencini $^{(a,b)}$, A. Torcini$^{(a,c)}$, D. Vergni
$^{(a,b)}$ and A. Vulpiani $^{(a,b)}$}
\address{$(a)$ Dipartimento di Fisica, Universit\'a ``La Sapienza'',
Piazzale Aldo Moro 2, I-00185 Roma, Italy} 
\address{$(b)$ Istituto Nazionale di Fisica della Materia, UdR and SMC 
Roma 1, Piazzale Aldo Moro 2, I-00185 Roma, Italy}
\address{$(c)$ Dipartimento di Energetica ``S. Stecco'',
Via S. Marta, 3 - I-50139 Firenze, Italy}
\maketitle
\begin{abstract}
Front propagation in two dimensional steady and unsteady cellular
flows is investigated in the limit of very fast reaction and sharp
front, i.e., in the geometrical optics limit.  In the steady case, by
means of a simplified model, we provide an analytical approximation
for the front speed, $v_{\mbox{\scriptsize{f}}}$, as a function of the
stirring intensity, $U$, in good agreement with the numerical results
and, for large $U$, the behavior $v_{\mbox{\scriptsize{f}}}\sim
U/\log(U)$ is predicted.  
The large scale of the velocity field mainly rules the front speed behavior
even in the presence of smaller scales.
In the unsteady (time-periodic) case, the
front speed displays a phase-locking on the flow frequency and, albeit
the Lagrangian dynamics is chaotic, chaos in front dynamics only
survives for a transient. Asymptotically the front evolves
periodically and chaos manifests only in the spatially wrinkled
structure of the front.
\end{abstract}
\pacs{PACS numbers:  47.70.Fw, 05.45.-a}
\keywords{Laminar Reacting Flows, Chaotic Advection}
\begin{multicols}{2}
\section{Introduction}
\label{sec:1}
Front propagation in fluid flows is relevant to many fields of
sciences and technology ranging from marine ecology \cite{bio,Abra2} to
chemistry \cite{chem,Epst} and combustion technology \cite{combustion}.  A
complete description of the problem would require to consider the
coupled evolution of reactants and velocity field, including the
modification of the advecting field induced by the reaction.
In general, this is a very difficult task \cite{A}.  
Here we consider a simplified, but still physically
significant, problem by neglecting the influence of the reactants
on the velocity field. This amounts to consider the reaction as 
a constant-density process. Aqueous auto-catalytic
reactions, and gaseous combustion with a large flow intensity but
sufficiently low values of gas expansion across the flame are
important examples of chemical-physical systems for which 
this approximations is appropriate \cite{Ronney}.

In the simplest model a scalar field, $\theta({\bf
r},t)$, which represents the fractional concentrations of the
reaction's products (i.e., $\theta=1$ indicates inert material,
$\theta=0$ fresh one and $0<\theta<1$ means that fresh material
coexists with products), evolves according to the
advection-reaction-diffusion equation \cite{Peter,Xin}:
\begin{equation}
\partial_t \theta + {\bf u}\cdot{\boldmath{{\mbox{$\nabla$}}}}\theta =
D_0 \Delta \theta + {1 \over \tau}f(\theta) \;,
\label{eq:rad} 
\end{equation}
where $D_0$ is the molecular diffusivity, and ${\bf u}$ is a given
incompressible (${\boldmath{{\mbox{$\nabla$}}}} \cdot {\bf u}=0$)
velocity field. $f(\theta)$ is the production term and $\tau$ its time
scale.

Two limiting cases of Eq.~(\ref{eq:rad}) have been previously studied
in detail: $f(\theta)\equiv 0$ and ${\bf u} \equiv {\bf 0}$. In the
former, the equation for a passive scalar is recovered (for a review
see Ref.~\cite{MK99}).  The latter corresponds to the
reaction-diffusion equation, which has gathered much attention since
the seminal works of Fisher and Kolmogorov-Petrovsky-Piskunov
(FKPP)\cite{FKPP37,Fischer} (see also Ref.~\cite{Xin} and references therein).

Eq.~(\ref{eq:rad}) can be studied for different geometries and boundary
conditions. For instance, one can consider an infinite strip in the
horizontal direction with a reservoir of fresh material on the right
and inert products on the left, and periodic boundary conditions along
the transverse direction.  With this geometry a front of inert
material (stable phase) propagates from left to right.  If the medium
is at rest with the FKPP production term, $f(\theta)=\theta
(1-\theta)$, the front propagates with an asymptotic speed and
thickness given by \cite{Xin,FKPP37,Fischer}
\begin{equation}
v_0=2\sqrt{{D_0 \over \tau}}\,,
\qquad \xi= c \sqrt{{D_0 \tau}}\,,
\label{eq:vfkpp}
\end{equation} 
where $c$ is a constant depending on the definition adopted for $\xi$.
This result is valid whenever $f(\theta)$ is a convex function
($f^{''}<0$) with $f(0)=f(1)=0$ and $f^{'}(0)=1$.  For non-convex
$f(\theta)$ only  upper and lower bounds for the front
speed can be provided \cite{Xin}.

A more interesting physical situation is when the velocity field is
non zero.  In this case, generally,  the front propagates
with an average limiting speed, $v_{\mbox{\scriptsize f}}$, enhanced
with respect to the fluid at rest ($v_{\mbox{\scriptsize f}}>v_0$).
For very slow reaction, by means of homogenization
techniques~\cite{MK99}, one can show that the front speed behaves as
in Eq.~(\ref{eq:vfkpp}) with $D_0$ replaced by a renormalized
diffusion coefficient, $D_{\mbox{\scriptsize {eff}}}$, the so-called
eddy diffusivity (see Ref.~\cite{MK99} for an exhaustive review on the
determination of the eddy-diffusivity). 
In realistic systems, the reaction time scale is of the same
order or (more frequently) faster than the velocity time scale
(fast reaction), so that a simple renormalization of $D_0$ is
not sufficient to encompass the dynamical properties of the
system~\cite{spagnoli}. In some cases the front speed can be still
obtained by Eq.~(\ref{eq:vfkpp}) renormalizing not only the diffusion
constant, $D_0 \to D_{\mbox{\scriptsize {eff}}}$, but also the
reaction time $\tau \to \tau_{\mbox{\scriptsize {eff}}}$ \cite{abeletal}.
However, a general method to compute $v_{\mbox{\scriptsize f}}$ for a
generic velocity field does not exist.

Here we consider the limit of very fast reaction and very thin front,
i.e., the so-called geometrical optics regime \cite{Ronney}.
Formally, this corresponds to the limit $\tau\to 0$ and $D_0 \to 0$
maintaining the ratio $D_0/\tau$ constant \cite{KAW88}: from
(\ref{eq:vfkpp}) this means that $v_0$ is finite and $\xi\to 0$.  In
this regime the front is identified as a surface (a line in
$2d$), and the effect of the velocity field is to wrinkle the front
increasing its area (length in $2d$) and thereby its speed~\cite{Peter}. 

As to the velocity field we consider steady and unsteady cellular
flows (i.e., with closed streamlines) in two-dimensions
\cite{Const,Pomeau,BorMaj,Ashurst}.  Since coherent vortical
structures are typically present in real hydrodynamical systems,
cellular flows offer an idealized (but non-trivial) model to study the
effects of this kind of structures on front propagation. Real flows,
e.g., turbulent flows, are usually characterized by a very complex
temporal dynamics and spatial development of scales. In this respect a
steady cellular flow is too simple. Therefore, we also consider either
the presence of small scale spatial structures in the velocity field,
or the effects of time dependence, which induces a complex temporal
behavior for particle trajectories --Lagrangian chaos
\cite{lagrangian,fapalvu}.

We found that the front speed is mainly determined by the large scales
velocity characteristics, i.e., the effect of small scales is just to
renormalize the largest scale velocity intensity.  Concerning time
periodic cellular flows, we found that the front speed is not
significantly modified with respect to the steady case, although
rather subtle effects appear. Namely the front speed locks on the flow
frequency: a phenomenon which goes under the name of mode-locking (or
frequency-locking) \cite{jens,pikovsky}, and which has been already
found in some models of front propagation \cite{cml}.  Lagrangian
chaos is suppressed by the reaction and only survives for a
transient. However, asymptotically the spatial wrinkling
(``complexity'') of the front is enhanced with respect to the steady
case. We introduced a suitable observable to quantify the front
complexity.

The paper is organized as follows. In Section~\ref{sec:2} we discuss
the geometrical optics limit. Numerical results for steady cellular
flows with one and more scales are presented in Section~\ref{sec:3},
where we propose a simple model which well reproduces the numerical
results.  The effects of Lagrangian chaos and velocity phase-locking
in time dependent cellular flows are discussed in Section~\ref{sec:4}.
Final remarks are reported in Section~\ref{sec:5}.  The Appendices are
devoted to the numerical methods here employed and to a more detailed
treatment of the frequency-locking phenomenon.

\section{The geometrical optics limit}
\label{sec:2}
From a physical point of view the geometrical optics limit (in
combustion jargon, the flamelet regime) corresponds to situations in
which the reaction time scale and reaction zone thickness are much
faster and much smaller than the time and length scales of the flow,
respectively (e.g., in turbulent flows this means that the front
thickness is smaller than the Kolmogorov length scale
$\ell_{K}$, $\xi\ll\ell_{K}$) \cite{Peter}.

Being the front sharp, its dynamics can be described in terms of the
evolution of the surface (line in $2d$) which divides the inert
material ($\theta=1$) from the fresh one ($\theta=0$). In this limit
the problem can be formulated in terms of the evolution of a scalar
field, $G({\bf r},t)$, where the iso-line (in 2D) $G({\bf r},t)=0$
represents the front: $G>0$ is the inert material and
$G<0$ is the fresh one. 
$G$ evolves according to the so-called $G$-equation
\cite{Peter,KAW88,Aldredge94,Aldredge96,Majda-Geq,McLaughlin}
\begin{equation}
{\partial G \over \partial t}+{\bf u}\cdot {\mbox{\boldmath $\nabla$}}
G=v_0 |{\mbox{\boldmath $\nabla$}}G|\,.
\label{eq:Geq}
\end{equation}
The analytical treatment of this
problem is not trivial, and even in relatively simple cases
(e.g., shear flows) numerical analysis is needed.

Recently Majda and collaborators \cite{Majda-Geq} pointed out
that there are situations in which the $G$-equation fails in
reproducing the front speed of the original
reaction-advection-diffusion model. Indeed, in some systems the
exact treatment of Eq.~(\ref{eq:rad}) in the limit $\tau\to 0$, $D_0
\to 0$ with $D_0/\tau=const$ does not lead to the same results of the
$G$-equation.  However, for the application we are interested in, the
study of the $G$-equation is physically relevant \cite{BorMaj}.

In the absence of stirring (${\bf u}={\bf 0}$) the front evolves
according to the Huygens principle, i.e., a point ${\bf x}$ belonging
to the front surface moves with a velocity ${\bf v}({\bf x})=v_0
\hat{\bf n}({\bf x})$, where $\hat{\bf n}({\bf x})$ is the
perpendicular direction to the front surface in ${\bf x}$. With open
boundary conditions, at large times the front surface is
asymptotically close to a sphere (circle in $2d$). However, the
preasymptotic behavior is mathematically non trivial \cite{Caglioti}
and  interesting in some technological problems.

In the presence of stirring (${\bf u}\neq {\bf 0}$) the problem is
much more difficult.  The first attempt to determine the front speed
in such a regime dates back to the $40$'s with the work of Damk\"oler
\cite{Peter} who suggested that, if the velocity field does not change
the local (bare) front speed, $v_0$, then the effective front speed is
proportional to the total front area divided by the cross-section flow
area.  In two-dimensional geometry this means that
\begin{equation}
v_{\mbox{\scriptsize f}}/v_0= L_{\mbox{\scriptsize f}}/L
\label{eq:relation}
\end{equation}
where $L$ is the transverse length, $v_{\mbox{\scriptsize f}}$ and
$L_{\mbox{\scriptsize f}}$ are the average front speed and length
respectively.

The average velocity and length are $v_{\mbox
{\scriptsize{f}}}=\langle v(t) \rangle$ and $L_{\mbox
{\scriptsize{f}}}=\langle {\cal L}(t) \rangle$, where $\langle\cdot
\rangle$ indicates a time average. The instantaneous front velocity,
$v(t)$, and length, ${\cal L}(t)$, can be defined as follows.  
The velocity is given by \cite{Const}
\begin{equation}
v(t)=\partial_t \left({1 \over L} \int_{0}^{L} \!{\rm d}y
\int_{-\infty}^{\infty}\!{\rm d}x \,  \theta(x,y;t)\right)\,\,.
\label{eq:vel}
\end{equation}
In order to define the instantaneous front length, ${\cal L}(t)$, 
we introduce the
variable $\sigma_\epsilon(x,y;t)$ which assumes the value $0$ if
$\theta$ is constant inside a circle of radius $\epsilon$ centered in
$(x,y)$, otherwise $\sigma_\epsilon(x,y;t) = 1$ (i.e.,
$\sigma_\epsilon(x,y;t) = 1$ only if the $\epsilon$-ball centered in
$(x,y)$ contains a portion of the front). The front length is then defined by
\begin{equation}
{\cal L}(t)= \lim_{\epsilon \to 0}{1\over \epsilon} 
\int_{-\infty}^{\infty}{\rm d}x \, 
\int_{0}^{L} {\rm d}y \, \sigma_\epsilon(x,y;t) \,\,.
\label{def:length}
\end{equation}

\section{Stationary Cellular Flow}
\label{sec:3}
We consider the following two-dimensional  cellular flow,
originally introduced in Ref.~\cite{Gollub} to mimic roll patterns in
Rayleigh-B\'ernard convection,
\begin{equation}
\left\{
\begin{array}{cc}  
u_x(x,y)= \!\!&U \sin\left({2\pi\over L}x\right)\cos\left({2\pi\over L}y\right)\\
\\
u_y(x,y)=\!\!&-U \cos\left({2\pi\over L}x\right)\sin\left({2\pi\over L}y\right)\,,
\end{array}
\right.
\label{eq:steadyflow}
\end{equation}  
where $U$ is the flow intensity, $L$ the roll size (in the following
for simplicity, $L=2\pi$). Periodic boundary conditions in the
transverse directions are assumed. By fixing $\theta=1$ for $x\to
-\infty$ and $\theta=0$ for $x\to \infty$ the front propagates from
left to right. It is natural to expect that the front speed can be
expressed according to the following relation
\cite{Ronney,Aldredge94,Aldredge96}
\begin{equation}
v_{\mbox{\scriptsize f}}=v_0 \psi\left({U\over v_0}\right)\,.
\label{eq:fun}
\end{equation}
Our aim is to investigate the dependence of $v_{\mbox{\scriptsize f}}$
on the flow intensity $U$.
\begin{figure} 
\epsfxsize=8truecm
\epsfysize=6truecm
\epsfbox{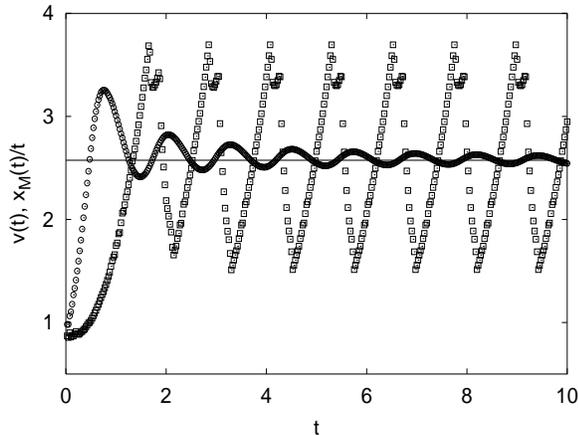}
\narrowtext
\caption{Front speed as a function of time,  measured in the standard 
way $v(t)$ (\ref{eq:vel}) ($\Box$) and as $x_M(t)/t$ (\ref{def:vel2})(o).  
The straight line is the average front speed $v_{\mbox{\scriptsize
 f}}$.  The system parameters are $U=4$, $v_0=1$ and $L=2\pi$.}
\label{fig:uno}
\end{figure}
Due to the spatial periodicity of the flow, after an initial
transient, the front propagates periodically in time (with period
$T$). In Fig.~\ref{fig:uno} a typical time series of the instantaneous
velocity $v(t)$ is reported: peaks occur when the front length is
maximal.  The front speed can be defined as the average
$v_{\mbox{\scriptsize f}}=\langle v(t)\rangle_T$ over a period. Since
the interface is sharp, we can easily track the farther edge of the
interface ($x_{M}(t),y_{M}(t)$), i.e., the rightmost point (in the
$x$-direction) for which $\theta(x_M,y_M;t)=1$. Thus we can 
define a velocity
\begin{equation}
   \tilde{v}_{\mbox{\scriptsize f}} = \lim_{t \to \infty} {x_M(t) \over t}\,,
   \label{def:vel2}
\end{equation}
which, since the front dynamics is periodic, is equivalent to the
standard definition (\ref{eq:vel}) (see Fig.~\ref{fig:uno}).

\begin{figure} 
\epsfxsize=8truecm
\epsfysize=4truecm
\epsfbox{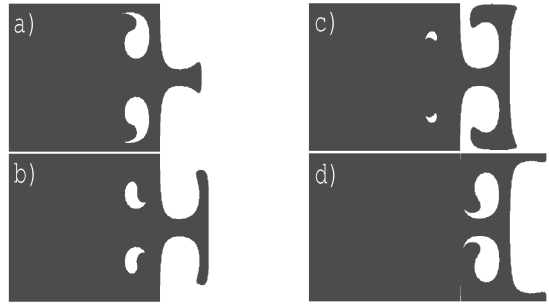}
\narrowtext
\vspace{0.3truecm}
\caption{Snapshot of the front shape with time step $T/8$
(from (a) to (d)), where $T$
is the period of the front dynamics, for $v_0 = 0.5$, $U=4.0$
and $L=2\pi$.}
\label{fig:due}
\end{figure}

In Fig.~\ref{fig:due} we show some snapshots of the front at different
times. It is possible to see cusps and pockets of unburnt material
(white) left behind the front edge, as firstly noticed by Ashurst and
Sivanshinsky \cite{AS91}.  At high field intensity a trail of pockets
is formed.

\begin{figure} 
\epsfxsize=8truecm
\epsfysize=6truecm
\epsfbox{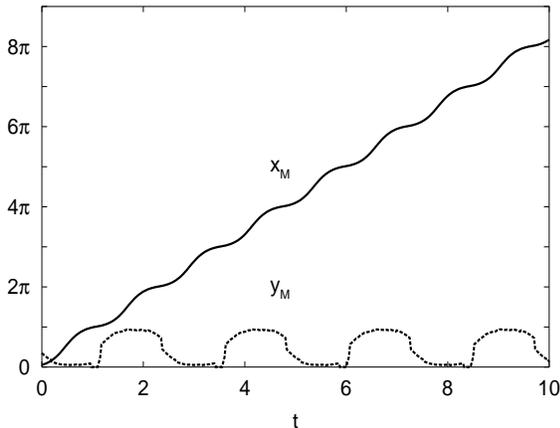}
\narrowtext
\caption{Time evolution of the edge point: $x_M(t)$ and $y_M(t)$.  The
simulation parameters are the same of Fig.~\ref{fig:uno}}
\label{fig:tre}
\end{figure}

We now revert to the behavior of $v_{\mbox{\scriptsize f}}$ as a
function of $U$.  As far as we know, apart from very simple shear
flows (for which $\psi({\cal U})={\cal U}+1$) \cite{sheartd}, there
are no general methods to compute $\psi({\cal U})$ from first
principles.  For the cellular flow under investigation, by mapping the
front dynamics onto a one dimensional problem, it is possible to
obtain an approximate expression for $\psi$ in good agreement with
numerical data.
The strategy is to devise an equation for the edge point evolution
(shown in Fig.~\ref{fig:tre}).  After the transient, in the cell
$[0,2\pi]\times[0,\pi]$, the point $(x_M(t),y_M(t))$ approximatively
moves in the right direction along the separatrices, so that $y_M(t)$
practically assumes values close to $0$ or $\pi$. Therefore, one can
reduce the edge dynamics to the following $1d$-problem
\begin{eqnarray}
{{\rm d}x_M \over {\rm d}t} = v_0 + U \beta |\sin(x_M)|\,.
\label{eq:simplemodel}
\end{eqnarray}
The second term of the r.h.s. mimics the horizontal component of the
velocity field, where $\beta$ takes into account the ``average''
effect of the dependence on the vertical coordinate, $y$. 
The periodicity of the r.h.s. of Eq.~(\ref{eq:simplemodel}) implies
that there exists a time $T_M$ such that $x_M(t+T_M)=x_M(t)+\pi$. 
Then, by solving (\ref{eq:simplemodel}) in
the interval $x_M\in [0,\pi]$ we computed $T_M$ and 
hence the front speed, $v_{\mbox{\scriptsize f}} = \pi / T_M$,
\begin{equation}
\psi_{\beta}({\cal U}) = {\pi {\sqrt{({\cal U} \beta)^2 - 1}} \over
2 \ln \left({\cal U}\beta+\sqrt{({\cal U} \beta)^2 - 1}\right)}
\label{eq:func}
\end{equation}
Notice that (\ref{eq:func}) is valid only for ${\cal U}\beta \geq 1$,
which is the regime we are interested in.
Now the problem is to estimate $\beta$. 

\begin{figure} 
\epsfxsize=8truecm
\epsfysize=6truecm
\epsfbox{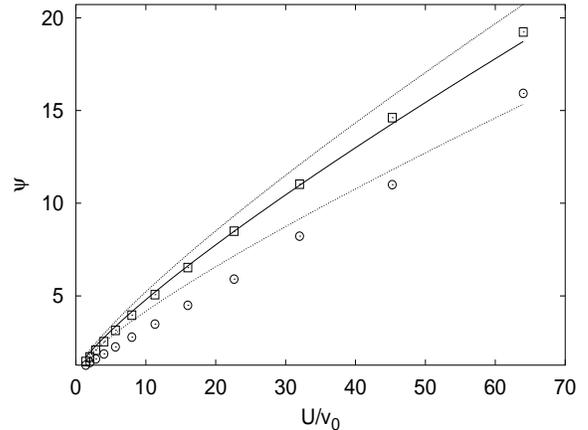}
\narrowtext
\caption{The measured $v_{\mbox{\scriptsize{f}}}/v_0$ as a function 
of $U/v_0$ ($\Box$), data obtained using the measured values in
Eq.~(\ref{eq:yakhot}) with $\alpha=2$ and $d=1/2$ (o) (see text),
the function $\psi(U/v_0)$ (\ref{eq:func}) for $\beta=0.875$ (solid line), 
and the upper and lower bounds, $\beta = 1$ in Eq.~(\ref{eq:func})
and Eq.~(\protect\ref{eq:oberman}), respectively (dotted lines).}
\label{fig:staz-vel}
\end{figure}

As stated previously the front evolution is periodic, so that
$y_M(t)$ is a periodic function of time with period $T_y$ commensurate
to $T_M$ (see Fig.~3); the front period $T$ should also be
commensurable to $T_M$ and $T_y$.  Therefore, for a specific value of
$U$ and $v_0$ we numerically identified the period $T_y$ and computed
$\beta$ as
\begin{equation}
\beta={1\over T_y}\int_0^{T_y} |\cos(y_M(t))| {\rm d}t\,,
\label{eq:beta}
\end{equation}
obtaining $\beta\approx 0.875$.  Using this value in (\ref{eq:func})
one has a remarkable agreement with the measured $v_{\mbox{\scriptsize
f}}$, see Fig.~\ref{fig:staz-vel} (the agreement is between $6\%$ and
$10\%$ for all the range of investigated values of $U$ and $v_0$).
Notice that, by definition, $\beta\!\leq \!1$. Therefore, with
$\beta\!=\!1$ in Eq.~(\ref{eq:func}) one obtains an upper bound for
the front speed.  Moreover, in Ref.~\cite{oberman2001}, a rigorous
lower bound has been provided:
\begin{equation}
v_{\mbox{\scriptsize f}} \geq U \, / \, \ln(1+U/v_0)\,.
\label{eq:oberman}
\end{equation}
It is worth remarking that, for large $U$, both the upper and lower
bounds give $v_{\mbox{\scriptsize f}}\sim\! U/\ln U$, 
which identify the asymptotic behaviour of the front speed.
Notwithstanding the considered cellular flow has only one spatial
scale, it is interesting to compare the results with the relation~\cite{KA92}
\begin{equation}
{v_{\mbox{\scriptsize f}} \over v_0} = 
\exp \left [ d \left ( {U_{\mbox{\scriptsize{rms}}} \over 
                        v_{\mbox{\scriptsize f}}} \right )^\alpha
     \right ]\,,
\label{eq:yakhot}
\end{equation}
originally proposed by Yakhot \cite{Yakhot88} and Shivanshinsky
\cite{shiva} for (multi-scale) turbulent flows, where
$U_{\mbox{\scriptsize{rms}}}$ is the turbulent intensity (i.e., the
root mean square velocity) and $\alpha=2$ and $d=1$ are two parameters
depending on the flow.  Indeed, Eq.~(\ref{eq:yakhot}) has been
frequently used in literature also for non turbulent flows and various
values of $\alpha$ have been reported \cite{Ronney,KA92}. In
particular, for the same cellular flow here studied,
Aldredge~\cite{Aldredge94,Aldredge96} compared his numerical results with
(\ref{eq:yakhot}) for $\alpha=2$ and $d=1/\sqrt{2}$ finding a fairly
good agreement (our data and Eq.~(\ref{eq:yakhot}) are shown in
Fig.~\ref{fig:staz-vel}).  Actually, the asymptotic behavior of
Eq.~(\ref{eq:yakhot}) is $v_{\mbox{\scriptsize f}} \sim U/(\ln
U)^{1/\alpha}$ which, by comparing with ours results, would suggest
$\alpha=1$.  In the following we investigate how small spatial scales
modify the dependence of $v_{\mbox{\scriptsize {f}}}$ on the flow
intensity.
\subsection{Effect of small scales}
\label{sec:3.1}
We consider now the following generalization of Eq.~(\ref{eq:steadyflow}):
\begin{eqnarray}
u_x(&&x,y)= U \sin\left(\frac{2 \pi}{L} x\right)\cos\left(\frac{2 \pi}{L}y\right)+ \nonumber
\\ 
&& \sum_{n=1}^{N} Q_n \sin\left(\frac{2 \pi}{L} k_n
x+\phi^x_{n}\right)\cos\left(\frac{2 \pi}{L} k_n y+\phi^{y}_{n}\right)
\nonumber 
\\ u_x(&&x,y)=- U \cos\left(\frac{2 \pi}{L}x\right)\sin\left(\frac{2 \pi}{L}y\right)-
\label{2scale}
\\
&& \sum_{n=1}^{N} Q_n \cos\left(\frac{2 \pi}{L} k_n
x+\phi^x_{n}\right)\sin\left(\frac{2 \pi}{L} k_n y+\phi^y_{n}\right)\,,
\nonumber
\end{eqnarray}
where $N$ is the number of different scales present in the flow,
$\{k_n\}$ are integers giving the ratio between the different spatial
scales, $Q_n$ is the velocity intensity at scale $\sim
1/k_n$, $\{\phi^{x}_n\,,\;\phi^y_n\}$ are  (time-independent) phase
differences.

\begin{figure} 
\centerline{\epsfig{figure=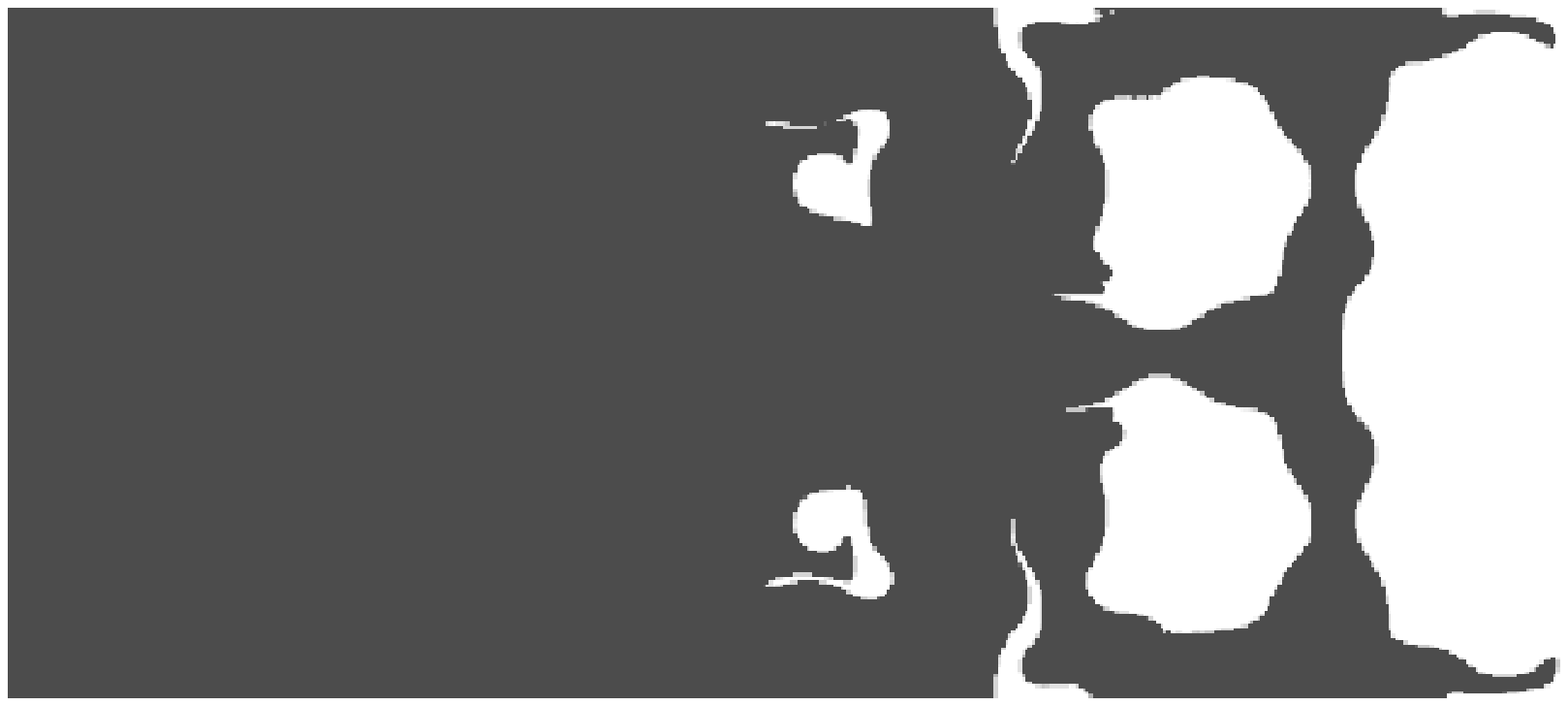,angle=270,width=8.0cm}}
\centerline{\epsfig{figure=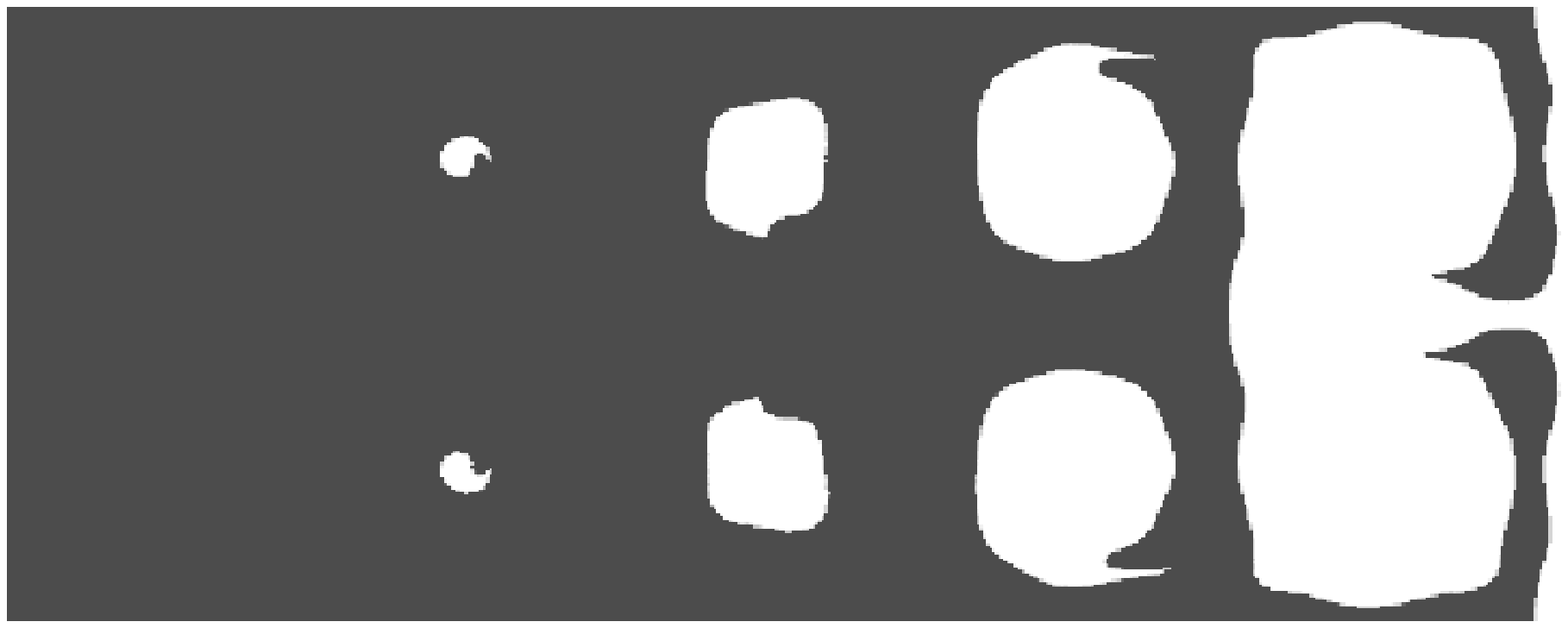,angle=270,width=8.0cm}}
\vspace{0.4truecm}
\caption{Images of the burnt area (black) for $U=6$, $Q=3$ (above) and
$U=10$, $Q=2$ (below) with $k=5$, where $v_0=0.5$. The different
scales are clearly visible.}
\label{fig:imagine}
\end{figure}

In Fig.~\ref{fig:imagine} we present two snapshots of the front for
different parameters values.  By comparing with Fig.~\ref{fig:due} it
is clear the presence of small structures in the front due to
smaller scales in ${\bf u}$.  

\begin{figure} 
\centerline{\hspace{-0.3truecm}\includegraphics[height=8.2cm,width=5.7cm,angle=270,draft=false]{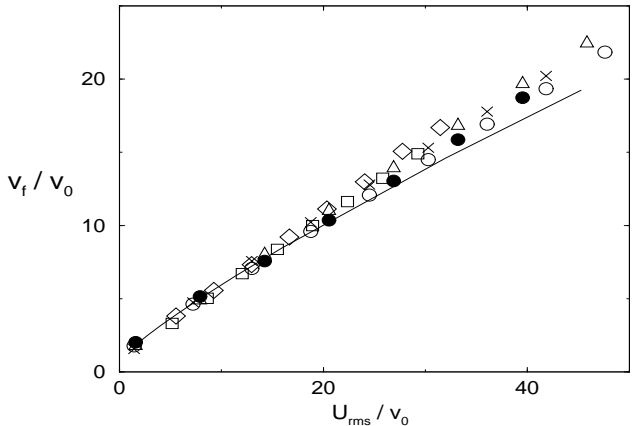}}
\vspace{0.2truecm}
\protect\caption{Front speeds $v_{\mbox{\scriptsize{f}}}/v_0$ as a
function of $U_{{\mbox{\scriptsize {rms}}}}/v_0$ for the 
flow~(\protect\ref{2scale}) for various $k_n$, $Q_n$-values 
and $v_0$ values.
Filled line is the one scale data with $v_0 = 1.0$. 
Two scales data, with $v_0 = 0.5$, are:
$Q_1=U/2$ and $k_1 = 3\, (\bigtriangleup)$,
$Q_1=U/5$ and $k_1 = 3\, (\times)$,
$Q_1=U/2$ and $k_1 = 5\, (\bullet)$,
$Q_1=U/5$ and $k_1 = 5\, (\circ)$.
For a turbulent caricature field we used $v_0 = 0.5$, 
$Q_n = U k_n^{-1/3}$, $k_n = 3^n$. The reported data refer to 
2 scales ($\Box$) and 3 scales ($\Diamond$), only.}
\label{fig:2scale}
\end{figure}

We computed $v_{\mbox{\scriptsize{f}}}$ for $N=1$ (two-scales
flow) and $N=2$ (three-scales flow) with
different values of $k_n$, $Q_n$ and random phases.  In the case
$N=2$, $Q_n=U\,k_n^{-1/3}$ has been chosen as a caricature of the
power spectrum of three dimensional turbulence.  The results, compared
with the one-scale flow ($Q_n\!=\!0$) are summarized in
Fig.~\ref{fig:2scale}, where $v_{\mbox{\scriptsize{f}}}$ is rescaled with $v_0$ and reported
as a function of $U_{{\mbox{\scriptsize {rms}}}}/v_0$
($U_{{\mbox{\scriptsize {rms}}}}\!=\!U\sqrt{(1+Q^2/U^2)/2}$, with
$Q^2\!=\!\sum_n Q_n^2$).  As one can see, they roughly collapse onto a
single curve together with the one scale results.
The physical information one can extract from the above result is that
the front speed is essentially determined by the large scale behavior
of the velocity field (i.e., the flow intensity). Indeed, as previously
observed~\cite{Ashurst} the absence of open channels can be more
important than the detailed multi-scale properties of the flow.
Let us conclude this section with some remarks.  For
the one scale flow in the steady case, we found that $v_{\mbox{\scriptsize{f}}} \sim
U/\ln U$ as the flow intensity $U$ (or $U_{{\mbox{\scriptsize
{rms}}}}$ equivalently) is very large. This
corresponds to the large $U$ limit of the Yakhot like formula
(\ref{eq:yakhot}) with $\alpha=1$.  Our results indicate that the main
effect of the small scales is to renormalize the stirring
intensity. Indeed  once
taken into account $U_{{\mbox{\scriptsize {rms}}}}$ all the curves
roughly collapse (Fig.~\ref{fig:2scale}).

However, because of the limited range of spatial scales here investigated
it is difficult to say something definitive on the front propagation
in a multi-scale velocity field. 

From Fig.s~\ref{fig:imagine} and \ref{fig:2scale} one can see that the
introduction of small scales on ${\mathbf u}$ causes a roughening of
the front shape at the same scales of the velocity field, and
has just a minor effect on the front speed. The problem of the effect
of small scales on the front wrinkling has been
studied also by Denet~\cite{Denet2} in the context of the flame
propagation equation (FPE) (which is similar to the Kardar Parisi
Zhang equation \cite{kpz}). In that model the small scales
influence on the large scale motion seems to be important. 
However, it is not possible to make a direct comparison between
the two results, because the FPE and Eq.~(\ref{eq:Geq}) generate
different dynamics.

\section{Non-stationary cellular flow}
\label{sec:4}
We now consider the problem of front propagation in the time
dependent cellular flow 
\begin{equation}
\!\!\left\{
\begin{array}{cc}
\!u_x(x,y,t)= \!\!& \;\;U \sin\left[{2\pi\over L}x+B \sin(\omega
               t)\right]\cos\left[{2\pi\over L}y\right]\\
\\
\!\!u_y(x,y,t)=\!\!&-U \cos\left[{2\pi\over L}x+B \sin(\omega
               t)\right]\sin\left[{2\pi\over L}y\right]
\end{array}
\right.
\label{eq:unsteadyflow}
\end{equation}  
where the term $B\sin(\omega t)$ mimics lateral oscillations of the
roll pattern, which are produced by the oscillatory instability
\cite{Gollub}.  As one can see the steady case (\ref{eq:steadyflow})
corresponds to $B\!=\!\omega\!=\!0$. When $B,\omega\neq 0$ chaotic Lagrangian
trajectories appear \cite{lagrangian,Gollub}. The presence of complex
particle trajectories constitutes a step toward more realistic flows.

We are mainly interested in addressing the two following issues.
First, since trajectories starting near the roll separatrices
typically have a positive Lyapunov exponent, it is natural to wonder
about the role of Lagrangian chaos on front propagation.  Second, from
previous works \cite{anomalo,solomon}, we know that for the time
dependent flow (\ref{eq:unsteadyflow}) the transport properties are
strongly enhanced, therefore it is worth to see if similar effects
are reflected also in the front speed.
\subsection{Effects of chaos: transient dynamics }
\label{sec:4.1}
A direct consequence of Lagrangian chaos is the exponential growth of
passive scalar gradients and material lines
\cite{lagrangian,fapalvu}: a (passive) material line of initial length
$\ell_0$ for large times grows as
\begin{equation}
\ell(t)\sim \ell_0 e^{\Lambda t} \,.
\label{eq:linegrowth}
\end{equation}
Where $\Lambda$ is the first generalized Lyapunov exponent,
$$
\Lambda=\lim_{t \to \infty} \lim_{|\delta {\bf r}(0)|\to 0} 
{1\over t}\ln \left\langle {|\delta {\bf r}(t)| \over |\delta 
{\bf r}(0)|}\right\rangle \,,
$$
which is in general larger than the maximum Lyapunov exponent
\cite{lagrangian,fapalvu}.  
\begin{figure} 
\centerline{\hspace{-0.truecm}\includegraphics[scale=0.30,draft=false]{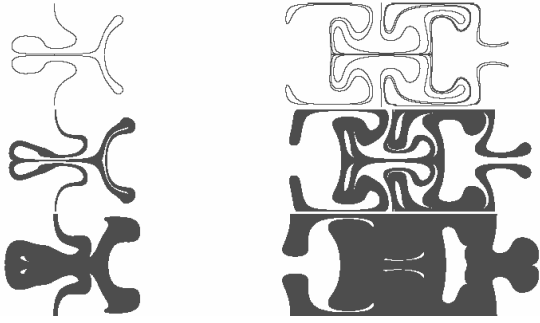}}
\vspace{0.4truecm}
\caption{Snapshots at two successive times, $t=3.6$ and $7.5$, of the
evolution of passive (top) and reactive line of material for two
values of $v_0$ (middle $v_0=0.7$ and bottom $v_0=2.1$) for $U=1.9$,
$B=1.1$ and $\omega=1.1 U$. The initial condition is a straight
vertical line.}
\label{fig:confronto}
\end{figure}
The average in the previous equation is taken
along the Lagrangian trajectories. In the presence of molecular
diffusivity, the exponential growth of $\ell(t)$ stops due to
diffusion \cite{lagrangian} and chaos is just a
transient \cite{transient}. 
\begin{figure} 
\centerline{\includegraphics[scale=0.7,draft=false]{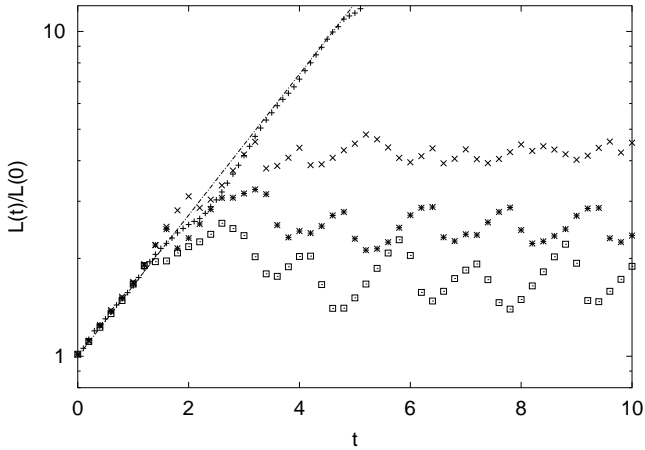}}
\vspace{0.4truecm}
\caption{
${\cal L}(t)/{\cal L}(0)$ as a function of time for $U\!=\!1.9$,
$B\!=\!1.1$ and $\omega\!=\!1.1 U$ for the passive ($+$) and reactive case: 
 from top $v_0=0.3\,\,(\times),\,\;
0.5\,\,(*),\,\; 0.7\,\,(\Box)$.  The straight line indicates the
curve $\exp(\Lambda t)$ with $\Lambda \approx 0.5$, which has been
directly measured. }
\label{fig:lfversust}
\end{figure}
For reacting scalars something very similar happens.  Let us compare
the evolution of material lines in the passive and reactive cases (see
Fig.~\ref{fig:confronto}).  While in the passive case structures on
smaller and smaller scales develop (due to stretching and folding), in
the reactive one after a number of folding events structures on
smaller scales are inhibited as a consequence of the Huygens
dynamics: the interface between the two phases merges.  This
phenomenon is responsible for the formation of {\it pockets}
\cite{Aldredge94,Aldredge96,AS91}.  Of course, ``merging'' is more and more
efficient as $v_0$ increases (compare the middle and lower pictures of
Fig.~\ref{fig:confronto}).

In Fig.~\ref{fig:lfversust} we show the time evolution of the line
length, ${\cal L}(t)$, as a function of $t$ for the passive and
reactive material at different values of $v_0$.  It is clear from the figure
that while at small times both the passive and reactive scalar lines
grow exponentially with a rate close to $\Lambda$, at large time
$t>t^*$ (where $t^*$ is a transient time depending on $v_0$) the
reacting ones stops due to merging. At stationarity, the front length
varies periodically with an average value depending on $v_0$.  A rough
argument to estimate $t^*$, is the following:
two initially separated part of the line (e.g., originally at distance
$\ell_0$) become closer and closer, roughly as $\sim \ell_0
\exp(-\Lambda t)$. When such a distance becomes of the order of $v_0
t$ the merging takes place, hence to leading order
\begin{equation}
t^* \propto \frac{1}{\Lambda} \ln \left({\Lambda \ell_o \over v_0}\right)\,.
\label{eq:tcut}
\end{equation}
In the asymptotic state ($t>t^*$) both the
spatial and temporal structures of the flow become periodic.
\subsection{Effects of chaos: asymptotic dynamics }
\label{sec:4.2.2}
\begin{figure} 
\centerline{\includegraphics[scale=0.7,draft=false]{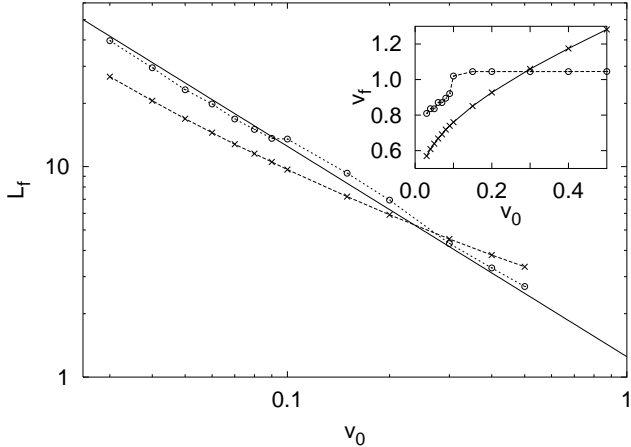}}
\vspace{0.4truecm}
\caption{The average front length $L_{\mbox{\scriptsize f}}$ as a
function of $v_0$ for the time dependent flow $(\circ)$, with $U=1.9$,
$B=1.1$, $\omega=1.1U$ and the time independent case $(\times)$ with
$U=1.9$. The straight line indicates the $1/v_0$ behavior.  In the
inset it is displayed $v_{\mbox{\scriptsize f}}$ versus $v_0$ for the
time dependent $(\Box)$ and for the time independent case.}
\label{fig:LFvsv0}
\end{figure}

Let us now switch to the effects of Lagrangian chaos on the asymptotic
dynamics of front propagation.  An immediate consequence of
Eq.~(\ref{eq:tcut}) is that the asymptotic front length
(\ref{def:length}) behaves as $L_{\mbox{\scriptsize f}} \sim
v_0^{-1}$ for enough small values of $v_0$. Indeed,
\begin{equation}
L_{\mbox{\scriptsize f}} \sim L e^{\Lambda t^*} 
                         \sim \frac{L^2\Lambda}{v_0} \,\,,
\label{eq:lfvsv0}
\end{equation}
which is in fairly good agreement with the simulations (see
Fig.~\ref{fig:LFvsv0}). It is interesting to note that in the steady
case a different scaling can be seen.  
For the sake of completeness, in the inset of Fig.~9 we display the
dependence of the front speed on $v_0$ for both the time-dependent and
time-independent flow.  For very small $v_0$, when chaos is effective
in enhancing the front length, one observe an
increasing of the front speed. At large values of $v_0$ the
time-independent flow has a larger front speed than the time-dependent
one. As we will see in the next subsection, this is a consequence of
the mode-locking of the dynamics, which maintains constant the value
of $v_{\mbox{\scriptsize f}}$.

It is worth remarking that even if the scaling (\ref{eq:lfvsv0}) holds
when chaos is present, in general it is not peculiar of chaotic flows.
For instance, for the shear flow ($u_x=U\sin(y)\,,\;u_y=0$) one has
$v_{\mbox{\scriptsize f}}=U+v_0$. On the other hand from
Eq.~(\ref{eq:relation}) $v_{\mbox{\scriptsize f}}\sim
L_{\mbox{\scriptsize f}} v_0$. Therefore, even if the shear flow is
not chaotic $L_{\mbox{\scriptsize f}}\sim 1/v_0$ for $U/v_0\gg 1$.
From the previous discussion, it seems that the front length
dependence on $v_0$ is not an unambiguous effect of chaos on the
asymptotic dynamics. But, comparing
Fig.~2 with Fig.~7 appears that the spatial ``complexity'' of the
front in the presence of Lagrangian chaos is higher than in its
absence.  
\begin{figure} 
\centerline{\includegraphics[scale=0.7,draft=false]{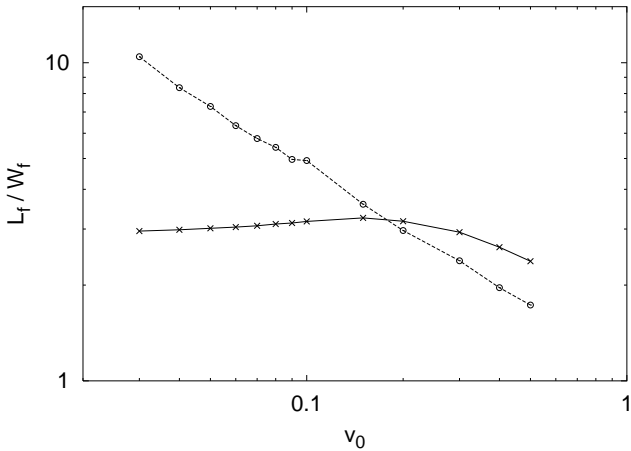}}
\vspace{0.4truecm}
\caption{$L_{\mbox{\scriptsize f}} /W_{\mbox{\scriptsize f}}$ as a
function of $v_0$ for the time dependent $(\circ)$ and independent
$(\times)$ cases for the same parameters of Fig.~9.}
\label{fig:vfvsv0}
\end{figure}
In the sequel we introduce an indicator to
quantitatively evaluate this qualitative observation.  

Let us call $W_{\mbox{\scriptsize f}}$ the size of the region in which
burnt and unburnt material coexist. Introducing a measure, $\mu(x)$, 
in that region we can defined $W_{\mbox{\scriptsize f}}$ 
as the standard deviation of $\mu(x)$~\cite{cml}:
\begin{equation}
\mu(x)= {|\partial_x \tilde{\theta}(x)|  \over \int {\rm d}x 
         |\partial_x \tilde{\theta}(x)|}\,,
\end{equation}
where $\tilde{\theta}(x)= {1 / L}\int_0^{L} \theta(x,y) {\rm d}y$, i.e.,
\begin{equation}
W_{\mbox{\scriptsize f}} = \left( \int x^2 \mu(x){\rm d}x -(\int  x \mu(x) 
{\rm d}x)^2\right)^{1/2} \,.
\end{equation} 

For a simple shear flow $W_{\mbox{\scriptsize f}}$ and
$L_{\mbox{\scriptsize f}}$ display the same kind of dependence on
$v_0$ (actually they are proportional).  In generic chaotic flows
there is an increasing of the front length, while chaotic mixing
induces a decrease of $W_{\mbox{\scriptsize f}}$.
This is indeed what one observes in
Fig.~10, where we show the ratio $L_{\mbox{\scriptsize f}}/
W_{\mbox{\scriptsize f}}$ both for the non-chaotic and the chaotic
flow. For the latter this ratio
diverges for very small $v_0$ values as a signature of chaos.  From a
physical point of view the ratio $L_{\mbox{\scriptsize{f}}}/W_{\mbox{\scriptsize{f}}}$ is an indicator of the
spatial complexity of the front. Indeed it indicates the degree of
wrinkling of the front with respect to the size of the region in
which the front is present.  Loosely speaking,  we can
say that the {\it temporal} complexity of Lagrangian trajectories
converts in the {\it spatial} complexity of the front.
\subsection{Front speed dependence on the frequency}
\label{sec:4.2}
\begin{figure} 
\centerline{\includegraphics
[scale=0.7,draft=false]{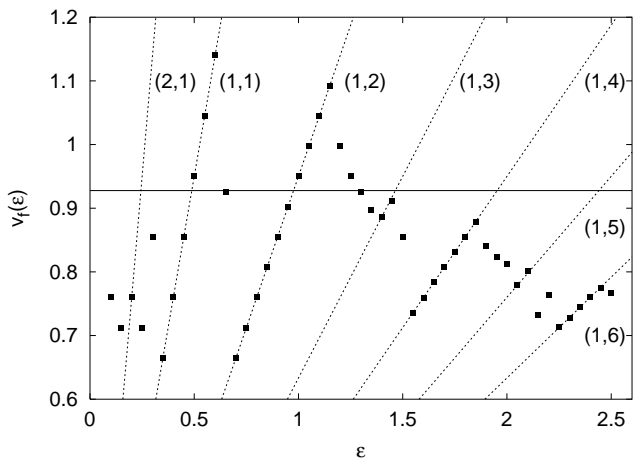}}
\vspace{0.4truecm}
\caption{$v_{\mbox{\scriptsize f}}(\epsilon)$ as a function of
$\epsilon=\omega/U$, for the flow (\ref{eq:unsteadyflow}) with
$U\!=\!1.9$, $v_0\!=\!0.2$ and $B\!=\!1.1$. The straight horizontal
line indicates the front speed for the steady case, $B\!=\!0$. The
dashed lines indicates the curves $U \epsilon N/M$ for different
$(N,M)$ integers.  }
\label{fig:vfsuvom0}
\end{figure}

For passive particles transport in the flow (\ref{eq:unsteadyflow}),
it has been found that the eddy diffusivity coefficient $D_{\mbox
{\scriptsize{eff}}}(\omega)$ displays a complex behavior with
resonant-like patterns \cite{anomalo,solomon}, in which the values of
$D_{\mbox {\scriptsize{eff}}}(\omega)$ are orders of magnitude larger
than the stationary-flow value, $D_{\mbox {\scriptsize{eff}}}(0)$. The
physical mechanism responsible for the resonances is related to the
interplay between the oscillation of the separatrices and the
circulation inside the cell. When circulation and oscillation
``synchronize'', a very efficient and coherent way of transferring
particles from one cell to the other takes place. Does it happen
something similar to the front speed in the reactive case?  In
Fig.~\ref{fig:vfsuvom0} we report $v_{\mbox{\scriptsize f}}
(\epsilon)$ as a function of $\epsilon=\omega/U$.  As one can see,
$v_{\mbox{\scriptsize f}}(\epsilon)$ varies both above and below the
time independent value, $v_{\mbox{\scriptsize f}}(0)$, and its range
of variability (about $30\%$ above or below the time independent value
$v_{\mbox{\scriptsize f}}(0)$) is very small compared with that of the
diffusion coefficient.  This  can be understood as a consequence of
the inequality $v_{\mbox{\scriptsize f}}\leq U+v_0$.  If one changes
the value of $v_0$ then the curve display a small shift up (increasing
$v_0$) or down (decreasing $v_0$) with some slight variations of the
values of the peaks. This behavior is very robust with respect to
changes of the flow parameters ($U$ and $B$).  Moreover,
$v_{\mbox{\scriptsize f}}(\epsilon)$ is a piece-wise linear function
of $\epsilon$ with slope given by $U$ multiplied by a rational
number. This phenomenon --called frequency
locking~\cite{jens,pikovsky}-- naturally arises in some non linear
oscillators which are periodically forced, when the oscillator
synchronizes its frequency to the one of the periodic forcing.

The frequency locking of the front speed can be understood with the
following argument.  At large times, $t>t^*$, the front is time
and space periodic. This means that after a time period $T$, the front
is rigidly translated in the $x$-direction by $S$, which is the
spatial period.  Due to the spatial periodicity of the flow
(\ref{eq:unsteadyflow}) $S= 2\pi N$ (where $N$ integer).  Now, as
confirmed by the simulations, one expects $T$ to be a multiple of the
oscillation period $T_o=2\pi/\omega$ so that: $T=M T_o$ (with $M$
integer). On the other hand, the front speed is nothing but $S/T$ so
that
\begin{equation}
 v_{\mbox{\scriptsize f}} = {S \over T} = {2\pi N  \over M
   T_o}= {N \over M} \omega={N \over M} U \epsilon\,,
\label{eq:locking}
\end{equation}
which is indeed the behavior we observed in Fig.~\ref{fig:vfsuvom0}.
Varying  $\omega$ the periods $S=2\pi M$ and $T=N T_o$
with a given $M$ and $N$ can loose their stability so that a new
couple of $N,M$ values is selected. This explains the presence of 
different linear behaviors. 
By generalizing the one
dimensional model (\ref{eq:simplemodel}) to the time dependent case
one can qualitatively reproduce the behavior of the front
speed dependence on the frequency (see Appendix~\ref{app:lock}).

It is interesting to compare the behavior of front propagation with
the time dependence (\ref{eq:unsteadyflow}) with previous studies that
considered a different time dependence, i.e., with $U \rightarrow U
\cos(\omega t)$ and $B=0$ in (\ref{eq:unsteadyflow}).  With this
choice the overall effect (as recognized by most of authors
\cite{AS91,Denet} see also \cite{flow-frequency}) is a depletion of the front wrinkling at increasing
the flow frequency. As a consequence, a strong bending of the front
speed with respect to the steady case has been observed. This
phenomenon has been quantitatively understood for the case of time
dependent shear flows (i.e., ${\bf u}=(U\sin(\omega t)\sin(y),0)$) by
Majda and collaborators \cite{sheartd}.  With the choice
(\ref{eq:unsteadyflow}) such a depletion is not observed
because of the enhanced transport properties of the flow
Fig.~\ref{fig:vfsuvom0}).

\section{Final remarks}
\label{sec:5}
In this paper we studied thin front propagation in steady and unsteady
cellular flows. In particular, we investigated the behavior of the front
speed and the front spatial structure at varying the system parameters.

As far as the one-scale steady case is concerned, we were able to give
a quantitative description of the front speed by means of a simple one
dimensional model.  For large flow intensity $U$ (or
$U_{\mbox{\scriptsize{rms}}}$ equivalently), the front speed behaves
as $v_{\mbox{\scriptsize{f}}}\sim
U_{\mbox{\scriptsize{rms}}}/\log(U_{\mbox{\scriptsize{rms}}})$, which
corresponds to the asymptotic behavior of the Yakhot-like formula
(\ref{eq:yakhot}) with $\alpha=1$. Moreover, small scales structures
have been added to the flow in order to study their effect on the
front speed.  Numerical simulations show that, once
$v_{\mbox{\scriptsize{f}}}$ is rescaled with $v_0$ and reported as a
function of $U_{{\mbox{\scriptsize {rms}}}}/v_0$, the results for the
one-scale flow and those for more than one-scale fairly collapse onto
a single curve. Therefore, the front speed is essentially determined by
the large scale behavior of the velocity field.

Small scales spatial structures may also be induced by Lagrangian
chaos. In this respect, on the basis of our results on the unsteady
cellular flow it is interesting to remark that the effect of chaos is
limited to a transient, in which the front behavior is close to the
passive scalar case.  Asymptotically, the reacting term induces a
drastic regularization on the front evolution, suppressing the effects
of small scales and Lagrangian chaos. Indeed the front propagates
periodically and displays a frequency locking phenomenon.

The only asymptotic effect of Lagrangian chaos is in the structure of
the front which is more and more wrinkled as $v_0$ approaches to
zero. On the contrary in the case of steady velocity fields (regular
Lagrangian motion) the degree of irregularity does not change with
$v_0$.  As an indicator of the spatial ``complexity'' of the front we
used the ratio between the front length and the width, $L_{\mbox{\scriptsize{f}}}/W_{\mbox{\scriptsize{f}}}$ which
is large (diverging as $v_0 \to 0$) for the unsteady case and is
roughly constant for the steady one.
\section{Acknowledgments}
We gratefully thank A.~Malagoli and A. Celani for discussions and
correspondences.  This work has been partially supported by the INFM
{\it Parallel Computing Initiative} and MURST (Cofinanziamento {\it
Fisica Statistica e Teoria della Materia Condensata}). M.C., D.V. and
A.V. acknowledge support from the INFM {\it Center for Statistical
Mechanics and Complexity} (SMC).

\begin{appendix}
\section{Numerical algorithm}

In numerical approaches one is forced to discretize both
space and time. 
We introduce a lattice of mesh size $\Delta x$ and $\Delta y$ 
(where we assume $\Delta x = \Delta y$) 
so that the scalar field is defined on the points 
${\mathbf x}_{n,m} = (n\Delta x,m\Delta y)$: 
$\theta_{n,m}(t)=\theta(n\Delta x, m\Delta y, t)$.\\
The time discretization implies a discretization
of the dynamics. Looking at the G-equation~(\ref{eq:Geq})
one immediately recognize two different terms:
the advection term ${\bf u}\cdot {\mbox{\boldmath $\nabla$}} G$,
accounting for the transport properties of the flow,
and the ``optical'' term $v_0 |{\mbox{\boldmath $\nabla$}}G|$,
which locally propagates the front in a direction perpendicular
to it with a bare velocity $v_0$.

Let us call ${\mathbf F}^{\Delta t}$ the Lagrangian propagator
for the discretized advection equation, $\partial_t {\bf x} =
 {\bf u}\cdot {\mbox{\boldmath $\nabla$}} G$.
Then, given the field at time $t$ one can computes
the field at time $t+\Delta t$ using a two steps algorithm:
\begin{itemize}
   \item [1)] using the Lagrangian propagator, ${\mathbf F}^{\Delta
   t}({\bf x})$, one evolves each point of the interface between burnt
   and unburnt region; 
  \item [2)] at each point of the evolved
   interface one constructs a circle of radius $v_0\,{\rm \Delta}t$,
   obtaining the new frontier as the envelope of the circles.
\end{itemize}

\begin{center}
\begin{figure}[htb]
\vspace{-1.3truecm}
\epsfig{figure=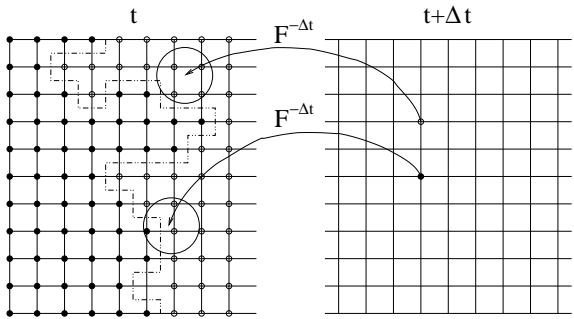,width=0.5\textwidth,angle=0}\hfill
\caption{Pictorial scheme of the numerical algorithm for the geometrical optics limit.}
\label{fig:nummetopt}
\end{figure}
\end{center}
The numerical algorithm can be easily implemented
using a time reversal procedure:
starting on a grid point, ${\bf x}_{n,m}$, of the scalar field
at time $t + \Delta t$ one applies the backward evolution 
obtaining the point ${\bf y} = {\mathbf F}^{-\Delta t} {\bf x}_{n,m}$ 
at time $t$. Around $y$ we construct the circle of radius 
$v_0\,{\rm \Delta}t$. If in this circle there is at least
one burnt point of the scalar field at time $t$, we fix 
$\theta({\bf x}_{n,m};t+{\rm \Delta}t) = 1$
otherwise $\theta({\bf x}_{n,m};t+{\rm \Delta}t) = 0$.

It is worth to note that one has to care about the radius of circle
$v_0\,{\rm \Delta}t$ that has to be larger than the grid-size 
$\Delta x$ (empirically we found that $v_0 \Delta t\geq 3-4 \Delta x$
it is enough to give good results).

Typical values used in our simulation are $\Delta t = 0.02$,
$\Delta x = 2\pi / 2048$. The backward Lagrangian integration
${\mathbf F}^{-\Delta t} ({\bf x})$ has been performed
with a $4$-th order Runge-Kutta algorithm.


\section{Front speed locking}
\label{app:lock}
Frequency Locking arises in many physical systems ranging from
Josephson-junction arrays to chemical reactions and non linear
oscillators \cite{jens,pikovsky,cml}. The basic mechanism is a
resonance effect between two oscillators or when an oscillator is
coupled with an external periodic forcing.  In the last case the
system synchronizes with the external forcing making its internal
frequency commensurable with the external one.  Almost all the systems
displaying frequency locking can be mapped to the damped forced
non linear oscillators \cite{jens}:
\begin{equation}
\alpha {d^2{\theta} \over dt^2}+\beta {d{\theta} \over dt}+\gamma
\sin(\theta)=\delta + \sigma \cos(\omega t)\,.
\label{equazione}
\end{equation}
The solution $\theta(t)$ is periodic and the frequency, i.e.,
the average angular velocity, turns out to be:
\begin{equation}
\langle {d{\theta} \over dt}\rangle= \lim_{t \to \infty} {\theta \over
t}={M\over N}\omega
\label{eq:lock}
\end{equation}
with $M,N$ integers. Moreover, if (\ref{eq:lock}) is realized for a
certain set of the parameters there always exist an entire interval 
around their values where (\ref{eq:lock}) holds with the same 
values of $M$ and $N$. This kind of behavior persists
also when $\alpha=0$ and for other kind of non linear terms 
(i.e., the third term of the l.h.s.).
An exhaustive description of such a phenomenon can be 
found in Refs.~\cite{pikovsky}.
\begin{figure} 
\centerline{\includegraphics
[scale=0.7,draft=false]{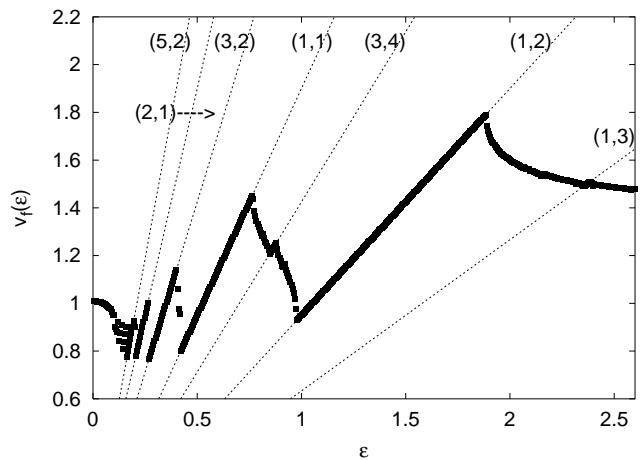}}
\vspace{0.2truecm}
\caption{$v_M(\epsilon)$ ad a function of $\epsilon=\omega/U$, for the
model (\ref{eq:modello}) with $U=1.9$, $v_0=0.2$ and $B=1.1$. The
dashed lines indicates the curves $U \epsilon N/M$ for different $N,M$
integers.  }
\label{fig:figmodello}
\end{figure}
Coming back to our system, we 
can generalize  the $1d$-model (\ref{eq:simplemodel}) to the
time dependent case:
\begin{equation}
{{\rm d}x_M(t)\over {\rm d}t} = v_0 + U |\sin(x_M+B\sin(\omega t))|\,.
\label{eq:modello}
\end{equation}
Note that in principle one should also take into account the dynamics
of $y_M$, but for the sake of simplicity we present just the
$y$-independent version of the model.  This model, although very
idealized, is able to reproduce behaviors qualitatively similar to the
ones observed in the simulations (compare Fig.~\ref{fig:figmodello})
 with Fig.~\ref{fig:vfsuvom0}). By
changing of variables $z(t)=x_M(t)+B\sin(\omega t)$
Eq.~(\ref{eq:modello})  reduces to
\begin{equation}
{{\rm d}z(t)\over {\rm d}t} = v_0 + U |\sin(z))|+B\omega \cos(\omega t) \,.
\label{eq:modello1}
\end{equation}
which corresponds to (\ref{equazione}) for $\alpha=0$, and for which frequency
locking has been studied in details.  It is interesting to quote a
recent work \cite{ultimo} which has studied the problem of locking in
a model very similar to (\ref{eq:modello1}) but in presence of
noise. They found that the locking phenomenon is rather robust under
the effect of noise and, moreover, it gives rise to resonances in the
diffusion coefficient.  All these results are qualitatively very
similar to the behavior of the system here studied.

\end{appendix}

\end{multicols}


\begin{thebibliography}{99}
\bibitem{bio} E.~R.~Abraham, 
``The generation of plankton patchiness by turbulent stirring'', 
Nature {\bf 391}, 577 (1998).
\bibitem{Abra2}
E.~R.~Abraham, C.~S~Law, P.~W.~Boyd PW, 
S.J~Lavender, M.~T~Maldonado and A.~R.~ Bowie, 
``Importance of stirring in the development 
  of an iron-fertilized phytoplankton bloom'', 
Nature {\bf 407}, 727 (2000).

\bibitem{chem} J.~Ross, S.~C.~M\"uller, and C.~Vidal, 
``Chemical waves'', Science {\bf 240}, 460 (1988).
\bibitem{Epst}
I.~R.~Epstein, ``The consequences of imperfect mixing in auto-catalytic 
chemical and biological systems'', Nature {\bf 374}, 231 (1995).

\bibitem{combustion} F.~A.~Williams, {\it Combustion Theory}
(Benjamin-Cummings, Menlo Park 1985).

\bibitem{A} S Malham and J. Xin, ``Global solutions to a reactive
Boussinesq system with front data on an infinite domain''
Comm. Math. Phys. {\bf 199}, 287 (1998).

\bibitem{Ronney} P.~D.~Ronney, 
``Some open issues in premixed turbulent combustion'', in {\it
Modeling in Combustion Science}, pp. 3-22, Eds. J. Buckmaster and
T. Takeno (Springer-Verlag Lecture Notes in Physics, 1994).

\bibitem{Peter} N.~Peters, 
{\it Turbulent combustion} (Cambridge University Press, 2000).

\bibitem {Xin} J. Xin, 
``Front Propagation in Heterogeneous Media'', 
SIAM Review {\bf 42}, 161 (2000). 

\bibitem{MK99} A.~J.~Majda, and P.~R.~Kramer, 
``Simplified models for turbulent diffusion: 
Theory, numerical modeling and physical phenomena'', 
Phys. Rep. {\bf 314}, 237 (1999).

\bibitem{FKPP37} A.~N.~Kolmogorov, I.~G.~Petrovskii, and
N.~S.~Piskunov, ``Study of the diffusion equation with growth of the
quantity of matter and its application to a biology problem'', Moscow
Univ. Bull. Math. {\bf 1}, 1 (1937).
\bibitem{Fischer} R.~A.~Fischer, ``The Wave of
Advance of Advantageous Genes'', Ann. Eugenics {\bf 7}, 355 (1937).

\bibitem{spagnoli} A.C.~Marti, F.~Sagues and J.M.~Sancho, 
``Front dynamics in turbulent media'', 
Phys. Fluids {\bf 9}, 3851 (1997).

\bibitem{abeletal}  M.~Abel, A.~Celani, D.~Vergni and A.~Vulpiani, 
``Front propagation in laminar flows'', 
Phys. Rev. E {\bf 64}, 046307 (2001).

\bibitem{KAW88} A.~R.~Kerstein, W.T.~Ashurst, F.~A.~Williams, 
``Field equation for interface propagation in 
an unsteady homogeneous flow field'', 
Phys. Rev. A {\bf 37}, 2728 (1988).

\bibitem{Const} P.~Constantin, A.~Kiselev, A.~Oberman and L.~Ryzhik,
``Bulk Burning Rate in Passive - Reactive Diffusion'', 
Arch. Rational Mechanics {\bf 154}, 53 (2000).

\bibitem{Pomeau} B.~Audoly, H.~Beresytcki and Y.~Pomeau,
``R\'eaction diffusion en \'ecoulement stationnaire rapide'',
C. R. Acad. Sci. {\bf 328}, S\'erie II b, 255 (2000).

\bibitem{BorMaj} A. Bourlioux and A.J. Majda, ``An elementary 
model for the validation of flamelet approximations in non-premixed
turbulent combustion'' Comb. Th. and Model.  {\bf 4}, 189 (2000).

\bibitem{Ashurst} W.T.~Ashurst, 
``Flame propagation through swirling eddies, a recursive pattern'', 
Comb. Sci. Tech. {\bf 92}, 87 (1993).

\bibitem{lagrangian} J.M.~Ottino, 
{\it The kinematics of mixing: stretching, chaos and transport}
(Cambridge University Press, 1989).
\bibitem{fapalvu} A. Crisanti, M. Falcioni,
G. Paladin and A. Vulpiani, ``Lagrangian Chaos: Transport, Mixing and
Diffusion in Fluids'', La Rivista del Nuovo Cimento {\bf 14}, 1
(1991).

\bibitem{jens} M.H.~Jensen, P.~Bak and T.~Bohr, ``Complete devil's
Staircase, Fractal Dimension, and Universality of Mode-Locking
Structure in the Circle Map'', Phys.Rev.Lett. {\bf 50}, 1637 (1983).

\bibitem{pikovsky}A.~Pikovsky, M.~Rosenblum, and J.~Kurths,
{\it Synchronization: A Universal Concept in Nonlinear Sciences},
(Cambridge University Press, 2001)

\bibitem{cml} R.~Carretero-Gonzalez, DK.~Arrowsmith and F.~Vivaldi,
``Mode-locking in coupled map lattices'' Physica D {\bf 103}, 381
(1997).

\bibitem{Aldredge94} R.~C.~Aldredge, ``The Scalar-Field Front
Propagation Equation and its Applications'', in {\it Modeling in
Combustion Science}, pp. 23-35, Eds., J. Buckmaster and T. Takeno
(Springer-Verlag Lecture Notes in Physics, 1994).
\bibitem{Aldredge96} R.~C.~Aldredge, ``Premixed Flame
Propagation in a High-Intensity, Large-Scale Vortical Flow'',
Comb. and Flame {\bf 106}, 29 (1996).

\bibitem{Majda-Geq} P.~F.~Embid, A.~J.~Majda and P.~E.~Souganidis,
``Comparison of turbulent flame speeds 
from complete averaging and the G-equation'', 
Phys. Fluids {\bf 7} (8), 2052 (1995).

\bibitem{McLaughlin} R.M.~ McLaughlin and J. Zhu, 
``The effect of finite front thickness 
on the enhanced speed of propagation'',
Comb. Sci. Tech. {\bf 129}, 89 (1997).

\bibitem{Caglioti} D.~Benedetto, E.~Caglioti and R.~Libero,
``Non-trapping sets and Huygens principle'' Rairo-Math Model Num, {\bf
33}, 517 (1999).

\bibitem{Gollub} T.H.~Solomon and J.P.~Gollub, 
``Chaotic particle transport in time-dependent 
Rayleigh-B\'enard convection'', 
Phys. Rev. A {\bf 38}, 6280 (1988).

\bibitem{AS91} W.T.~Ashurst and G.~I.~Shivanshinsky, 
``On flame propagation through periodic flow fields'', 
Comb. Sci. Tech. {\bf 80}, 159 (1991).

\bibitem{sheartd}B.~Khouider, A.~Bourlioux and A.J.~Majda, 
``Parameterizing the burning speed enhancement by small-scale periodic
flows: I. Unsteady shears, flame residence time and bending'',
Comb. Th. Model. {\bf 5} 295 (2001).

\bibitem{oberman2001} A.~Oberman, PhD Thesis Univ. of Chicago 2001.

\bibitem{KA92} A.~R.~Kerstein and W.T.~Ashurst, 
``Propagating rate of growing interfaces in stirred fluids'', 
Phys. Rev. Lett. {\bf 68}, 934 (1992).

\bibitem{Yakhot88} V.~Yakhot, 
``Propagation velocity of premixed turbulent flame'', 
Comb. Sci. Tech. {\bf 60}, 191 (1988). 

\bibitem{shiva} G.I.~Shivanshinsky, ``Cascade-renormalization theory
of turbulent flame speed'', Comb. Sci. and Tech. {\bf 62}, 77 (1988).

\bibitem{Denet2} B.~Denet, ``Are small scales of turbulence able 
to wrinkle a premixed flame at large scale?'', 
Comb. Theory and Modeling {\bf 2}, 167 (1998).

\bibitem{kpz} M.~Kardar, G.~Parisi, and Y.-C.~Zhang, Phys. Rev. Lett. {\bf 56},
889 (1986).

\bibitem{anomalo} P.~Castiglione, A.~Mazzino, P.~Muratore-Ginanneschi, 
A.~Vulpiani, ``On strong anomalous diffusion'', Physica D, {\bf 134},
75 (1999).  

\bibitem{solomon} T.~Solomon, A.~Lee, M.~Fogleman, 
``Resonant flights and transient super-diffusion in a time-periodic,
 two-dimensional flow'', Physica D {\bf 157} 40 (2001).

\bibitem{transient}A.K.~Pattanayak
``Characterizing the metastable balance between chaos and diffusion'', 
 Physica D, {\bf 148}, 1 (2001).

\bibitem{Denet}B.~Denet, ``Possible role of temporal correlations
in the bending of turbulent flame velocity'', Comb. Th. Model. {\bf
3}, 585 (1999).

\bibitem{flow-frequency} W.T.~Ashurst, ``Flow-frequency effect upon Huygens 
front propagation'', Comb. Th. Model. {\bf 4} 99  (2000).

\bibitem{ultimo} D.~Reguera, P.~Reimann, P.~H\"anggi and J.M.~Rub\`i,
``Interplay of frequency-synchronization with noise: current
resonances, giant diffusion and diffusion-crests'', Europhys. Lett. {\bf 57},
644 (2002).
\end{thebibliography}
\end{document}